\documentclass[pdflatex,sn-basic]{sn-jnl}

\usepackage{graphicx}%
\usepackage{multirow}%
\usepackage{amsmath,amssymb,amsfonts}%
\usepackage{amsthm}%
\usepackage{mathrsfs}%
\usepackage[title]{appendix}%
\usepackage{xcolor}%
\usepackage{textcomp}%
\usepackage{manyfoot}%
\usepackage{booktabs}%
\usepackage{algorithm}%
\usepackage{algorithmicx}%
\usepackage{algpseudocode}%
\usepackage{listings}%

\theoremstyle{thmstyleone}%
\theoremstyle{thmstyletwo}%

\theoremstyle{thmstylethree}%

\begin{document}

\title[A Bayesian mixed-effects model to evaluate the determinants of COVID-19 vaccine uptake in the US]{A Bayesian mixed-effects model to evaluate the determinants of COVID-19 vaccine uptake in the US}


\author{\fnm{Asim K.} \sur{Dey}}\email{a.dey@ttu.edu}

\affil{\orgdiv{Department of Mathematics and Statistics}, \orgname{Texas Tech University}, \orgaddress{\city{Lubbock}, \postcode{79409}, \state{Texas}, \country{United States}}}




 \abstract{\textbf{Objectives:} The COVID-19 pandemic has adversely affected US public health, resulting in over a hundred million cases and more than one million deaths. Vaccination is the key intervention against the COVID-19 pandemic. Multiple COVID-19 vaccines are now available for human use. However, several factors, including socio-demographic variables, impact the uptake of COVID-19 vaccines. This study aims to assess different socio-demographic and spatial factors that influence the acceptance of COVID-19 vaccines in the US.

 \textbf{Methods:} In this study, first, we apply hierarchical clustering to group US states according to their county-level COVID-19 vaccination rates. Second, we build a Bayesian mixed-effects model to assess different socio-demographic factors that influence the acceptance of COVID-19 vaccines in the US. 

 \textbf{Results:} 
 The hierarchical clustering algorithm of US states finds three main cluster regions: (i) the Northeast, (ii) most of the South, Midwest (except the Upper Midwest), and Mountain West, and (iii) the Pacific, the Upper Midwest, Florida, and the rest. 
 The results of the Bayesian mixed-effects model show that education and high household income significantly increase COVID-19 vaccine uptake. African American and Asian individuals have higher probabilities of accepting vaccines compared to white individuals. There is no significant difference between males and females in taking COVID-19 vaccines.

\textbf{Conclusion:} The socio-demographic and spatial factors, e.g., race, gender, higher education, household income, and geographic region, influence people’s COVID-19 vaccine acceptance in the US.

}




\maketitle

\section{Introduction}\label{sec1}

COVID-19 has had a catastrophic effect on the US, resulting in over 111 million cases and more than 1.2 million deaths~\citep{Worldometer}. 
The COVID-19 pandemic also profoundly impacts the U.S. economy, triggering a sharp recession in 2020 due to widespread lockdowns and business closures~\citep{federalreserve2020,Dey2022Crudeoil_COVID19,DEY2022126423}. Vaccination is one of the most successful and cost-effective interventions against COVID-19. About 230 million people in the US had been fully vaccinated as of May 2024, which constitutes 70\% of the US population, and at least 270 million people or 81\% of the population have received at least one dose~\citep{CDC,USA_FACTS}. However,  COVID-19 vaccine uptake is influenced by different factors~\citep{CDC_2_2021}.

Several studies have been conducted to evaluate different socio-demographic and spatial factors, e.g., race, gender, higher education, household income, and geographic region,  that determine people’s COVID-19 vaccine acceptance. \cite{Terry_2022} conduct a systematic review and meta-analysis for sociodemographic factors associated with COVID-19 vaccine intentions.
\cite{Viswanath_2021} examine the individual, communication, and social determinants associated with COVID-19 vaccine uptake. \cite{MOON2023102200} study the determinants of COVID-19 vaccine hesitancy in California, US. \cite{Hechter_2024} evaluate COVID-19 vaccination coverage among people with HIV. \cite{MALIK2020100495} conduct an online survey of the U.S. adult population and evaluate different demographic and geographical determinants of COVID-19 vaccine acceptance. \cite{Dong_2024} study U.S. county level determinants of COVID-19 vaccination rates. A number of other studies have also investigated COVID‑19 vaccine hesitancy in the US~\citep{Nguyen_2022,Gerretsen_2021,Mohandes_2021,Wang_2021,Abrams_2020,Stijven_2022,Roy_2023}.

However, these existing studies often conduct descriptive analyses of the data and typically pay limited attention to the uncertainties of the coefficients of the factors of the COVID‑19 vaccine acceptance. This study uses a Bayesian mixed-effects model and thoroughly evaluates different socio-demographic and spatial factors, e.g., gender, race, education, household income, and geographic regions (e.g. US states), that influence COVID-19 vaccine acceptance. In particular, we focus on the probabilistic inference of the determinants and their uncertainty quantification. The mixed-effects model combines fixed effects (e.g.,  gender, race, and education) and random effects (e.g., US states), and applies MCMC-based Bayesian estimation~\citep{Gelman2008,pinheiro2009mixed,Neal_2011,Betancourt_2013}. We build the Bayesian mixed-effects model in \textit{PyStan} environment~\citep{stan_development_team_stan_2012}. 


 The rest of the paper is organized as follows. Section~\ref{sec:Data}  describes the data and variables. The proposed mixed-effect model and MCMC-based Bayesian estimation are described in Section~\ref{sec:Model}. Section~\ref{sec:Results} presents findings and a discussion of the results. Finally, Section~\ref{sec:Conclusion} provides a conclusion.

\section{Data}
\label{sec:Data}

We apply two datasets: the Household Pulse Survey (HPS) data and the US county levels COVID-19 vaccination dataset. HPS data published by the US Census Bureau~\citep{Pulse_Survey}. The county levels COVID-19 vaccination rate data are obtained from the Centers for Disease Control and Prevention (CDC)~\citep{CDC}.

The HPS is a biweekly cross-sectional online survey that measures how emergent social and economic issues are impacting households across the US. In this study, we use the HPS Phase 3.7 dataset, which contains data between December 9, 2022 and December 19, 2022. The data set comprises information about the interviewed person's gender, race, education status, household income, state, and whether received the COVID-19 vaccine. The dataset contains individuals from all the US states except Alaska and Hawaii. We define the focus variables and their categories in Table~\ref{t:Variables}.

\begin{table*}[!ht]
 \centering
 \caption{Variables.}
\label{t:Variables}
\begin{tabular}{llcc}
  \hline
      &  &   \\ [-10pt]
Variable  & Category  \\
 \hline
COVID - 19 vaccine receive & (i) Yes, and (ii) No\\
Education & (i) High school graduate or less, (ii) Associate's Degree, \\
& (iii) Bachelor’s degree, and (iv) Graduate degree \\
Race & (i) Black, (ii) White, (iii) Asian, and (iv) Others\\
Income & (i) Less than \$35,000, (ii) \$35,000 to  \$74,999, \\
& (iii) \$75,000 to \$149,999, and  (iv) \$150,000 or above \\
Gender & (i) Female, and (ii) Male& & \\
\hline 
\end{tabular}
\end{table*}

The US county levels COVID-19 vaccination dataset consists of information about the COVID-19 vaccination series complete for the age group 18 and above (\%) for 3193 counties in the USA.  The dataset contains COVID-19 vaccination information until December 28, 2022, and is available for all the US states except Alaska and Hawaii.

\section{Methodology}
\label{sec:Model}

\subsection{Mixed Effect Model}

Let $Y_i$ denote the COVID-19 vaccines acceptance for the $i$th individual, $i=1, 2, \cdots, n$. We assume $Y_i$ follows a Bernoulli random variable with probability distribution:
\begin{eqnarray}
Y_i =
\begin{cases}
1~ \text{(Received vaccines)}, & \text{with probability $\pi_i$ },\\
0~ \text{(Not received vaccines)}, & \text{with probability $1- \pi_i$ }.
\end{cases}       
\end{eqnarray}
That is, $Y_i \sim$ Bernoulli$(\pi_i)$. We evaluate the individual-level and state-level COVID-19 vaccine acceptance in  the US based on a \textit{multilevel logistic regression} framework as follows:
\begin{eqnarray}
\label{Eq:MEM}
\ln\left(\frac{\pi_i}{1-\pi_i}\right) = \underbrace{\textbf{X}'_i {\beta}}_{Fixed \ effects} + 
\underbrace{\alpha_{j[i]}, }_{Random \ effects}
\end{eqnarray}

where the expected response is obtained by the nonlinear logistic response function as
 \begin{equation}\nonumber
\text{E}(Y_i)= \text{P}(Y_i = 1)= \pi_i = \frac{\exp \left(\textbf{X}'_i {\beta} + \alpha_{j[i]}\right)} {1+\exp \left(\textbf{X}'_i {\beta} + \alpha_{j[i]}\right)}, 
\end{equation}

$j=1, 2, \cdots, p_s$, $p_s$ is the total number of states. In this study, input variables include the state index $j[i]$, and categorical variables education, race, household income, and gender. We define the $\textbf{X}'_i {\beta}$ as
  \begin{eqnarray}
\textbf{X}'_i {\beta}=\beta_0 + \underbrace{\beta_1 X_{1i}  + \beta_2 X_{2i} + \beta_3 X_{3i}}_{Education} + \underbrace{\beta_4 X_{4i} + \beta_5 X_{5i} + \beta_6 X_{6i}}_{Race} \\ \nonumber
+ \underbrace{\beta_7 X_{7i} + \beta_8 X_{8i} + \beta_9 X_{9i}}_{Income} + \underbrace{\beta_{10} X_{10i}}_{Gender}.
\end{eqnarray}

Here, intercepts vary by US states, that is, Equation~\ref{Eq:MEM} builds a separate model with different intercepts within each US state. In this multilevel logistic regression setting, the model has its own matrix of predictors in each state level~\citep{gelman2007data,wood2017generalized}.

\subsection{Estimation}
\label{Sec:Bayesian}
 We assign weakly informative prior distributions to the model parameters.  The regression parameters  are specified as $$\beta_k \sim \mathcal{N}(0, \sigma_k^2)$$ for $k=1,2, \cdots, p_x$, where $p_x$ is the total number of regression coefficients~\cite{stan_development_team_stan_2012,Gelman2008}. 
We also assign the distribution of the random effects as $$\alpha_j \sim \mathcal{N}(0, \sigma_j^2),$$ where $j=1,2, \cdots, p_s$.

Once the priors are defined, it is straightforward to simulate from the posterior density of the Bayesian logistic model by Markov chain Monte Carlo (MCMC) using stan software~\citep{stan_development_team_stan_2012}. Here we use Hamiltonian Monte Carlo (HMC)~\citep{Neal_2011,Betancourt_2013,betancourt2018}. 

We interpret the estimated regression coefficient $\beta_k$ based on the odds ratio, which is $\exp(\beta_k)$, $k=1,2, \cdots, p_x$. The odds ratio can be interpreted as the estimated increase in the probability of success associated with a one-unit change in the value of the predictor variable. In general, the estimated odds ratio associated with a change of $d$ units in the predictor variable is $\exp(d*\beta_k)$~\citep{kutner2005applied,gelman2007data, pinheiro2009mixed}.


\section{Results}
 \label{sec:Results}

\subsection{Descriptive Analysis}

\begin{figure*}[!ht]
         \centering
       \includegraphics[width=0.99\textwidth]{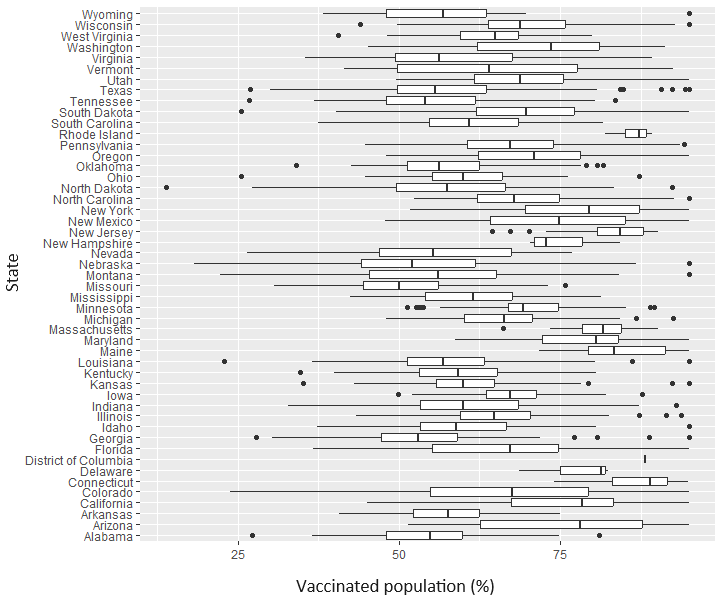}
       \caption{Boxplots of the county-level percentage of the fully vaccinated adult population in different US states.}
       \label{fig:Series_Complete_18PlusPop_Pct}
\end{figure*}

The COVID-19 vaccination dataset contains US county-level vaccination information. We conduct a summary analysis of the COVID-19 vaccination rate in different US states. Figure~\ref{fig:Series_Complete_18PlusPop_Pct} represents the boxplot of the percentage of  COVID-19 vaccine uptakes in state-wise counties. The boxplot gives us some available insights about the US state-wise vaccination acceptance. In particular, we find that the average percentages of the vaccinated adult population in most of the states are approximately between 50\% and 70\%. There are some states, e.g., Rhode Island, New Jersey, Maine, and Delaware, which have more than 80\% vaccination rate. In some states, the variability in county-level vaccination is very large. For example, in North Dakota, Nevada, Nebraska, Indiana, Florida, and Colorado, there are some counties in which the vaccination rate is very small, on the other hand, there are some counties in which the vaccination rate is very large.

\begin{figure*}[!ht]
         \centering
       \includegraphics[width=0.50\textwidth]{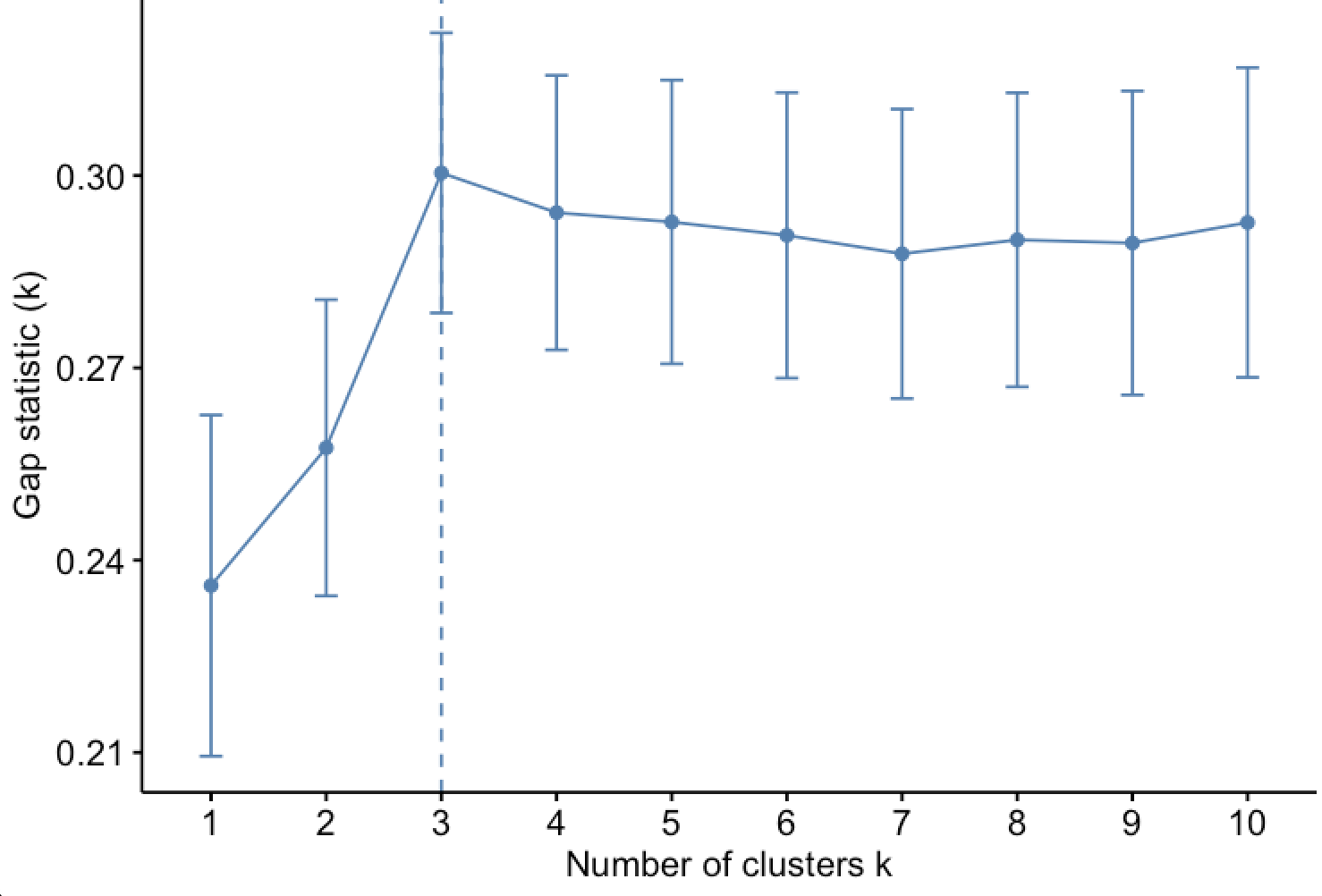}
       \caption{Selection of the optimal number of clusters of the US states for the fully vaccinated adult population.}
       \label{fig:HC0}
\end{figure*}

Now, to group US states according to their county-level COVID-19 vaccination rates, we apply a standard \textit{hierarchical clustering} algorithm~\citep{David_2001,Ghosal2019ASR,Benjamin_2023} to the US county-level COVID-19 vaccination rates. To determine the number of clusters $k$, we use the \textit{gap statistic}. The Gap Statistic finds the optimal number of clusters in the dataset by comparing how much the data within each cluster varies for different numbers of clusters $k$~\citep{Tibshirani_2001,Mojgan_2011}. We select the optimal number of clusters as 3 using the gap statistic (see Figure~\ref{fig:HC0}).

\begin{figure*}[!ht]
         \centering
       \includegraphics[width=0.54\textwidth]{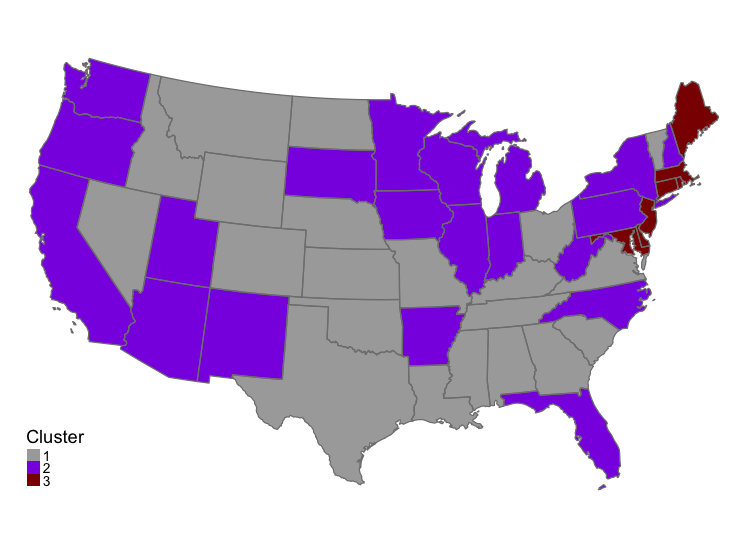}
       \includegraphics[width=0.44\textwidth]{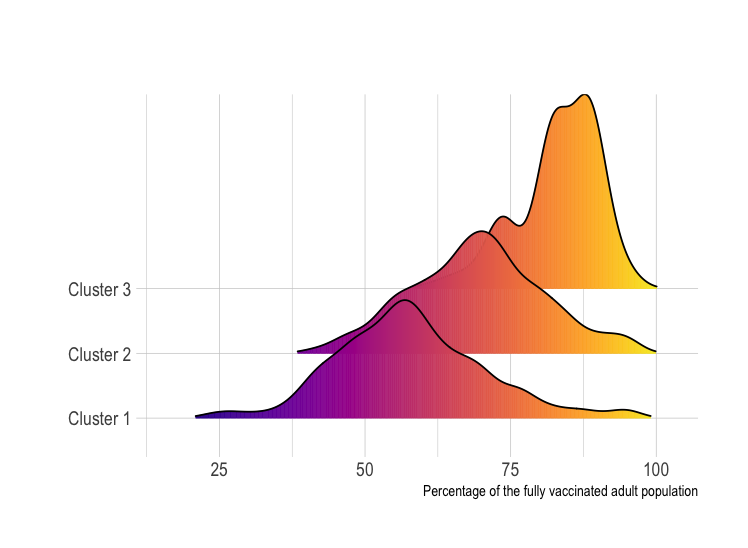}
       \caption{Clusters of the US states according to the fully vaccinated adult population (left panel). Density curves of the fully vaccinated adult population in different clusters of US states (right panel).}
       \label{fig:HC}
\end{figure*}

The left panel of Figure~\ref{fig:HC} shows the hierarchical clusters and their maps for the US COVID-19 vaccine acceptance. The clusters of different US states are represented by different colors. The clustering reveals that there are three main groups of states. That is, there are three obvious spatial differences in COVID-19 vaccinations in the US. The right panel of Figure~\ref{fig:HC} shows the probability densities of the fully vaccinated adult population in different clusters. We find that Cluster 3 contains states with the highest vaccine uptake, where the cluster average of the percentage of fully vaccinated adult population is 83.217 with a standard deviation of 6.809. Cluster 1 contains states with the lowest vaccine uptake, where the cluster average is 57.778 with a standard deviation of 12.839. Cluster 2 has moderate vaccine uptakes with an average of 69.223 and a standard deviation of 11.261.

Most of the Northeastern states, e.g., Maine, the District of Columbia (Washington, D.C.), Rhode Island, Maryland, Delaware, Massachusetts, Connecticut, and New Jersey, which have very high vaccination rates (See Figure~\ref{fig:Series_Complete_18PlusPop_Pct}) form one cluster (Cluster~1). Most of the South, Midwest (except the Upper Midwest) and Mountain West form one cluster (Cluster~1) with the lowest vaccine acceptance. The Pacific region, the Upper Midwest, and the rest create Cluster~2 with a medium vaccination rate.

\subsection{Mixed Effect Model Analysis}
 \label{sec:Results_2}

The parameters of the multilevel logistic regression model, Eq.~\ref{Eq:MEM}, are obtained based on the Bayesian methodology described in Section~\ref{Sec:Bayesian}. Here, MCMC is done based on 10000 iterations and 2 chains. Table~\ref{t:parameter} presents the estimated model parameters with 95\% credible intervals. Table~\ref{t:parameter} also gives the odds ratios of the categories of the variables compared to a base category.

\begin{table*}[!ht]
 \centering
 \caption{Estimated parameters with 95\% credible intervals (CI) and corresponding exponentiated coefficient (i.e., odds ratios).}
\label{t:parameter}
\begin{tabular}{llccccc}
  \hline
      &  & &  \\ [-10pt]
Variable  & Parameter &  Estimate (95\%  CI) &  Odds ratio   \\
 \hline
intercept & $\beta_0$     &  -0.74 (-1.32, -0.13)   &    \\ 
 
& & & & \\
Education: & & & \\

 Associate's Degree vs. High school graduate or less  & $\beta_1$  & 0.45 (0.35, 0.55) &  1.568 \\ 
 Bachelor's degree vs. High school graduate or less & $\beta_2$   & 1.24 (1.13, 1.35) & 3.456 \\ 
 Graduate degree  vs. High school graduate or less & $\beta_3$  & 1.79 (1.67,  1.92) & 5.989 \\ 

& & & & \\
Race: & & & \\
Black vs. White & $\beta_4$  &  0.39  (0.25,  0.53) & 1.477 \\ 
Asian vs. White & $\beta_5$  &  0.91 (0.66, 1.17) & 2.484 \\ 
Others vs. White & $\beta_6$   &   -0.3 (-0.46, -0.15)  & 0.741 \\ 

& & & & \\
Income: & & & \\
\$35,000 to  \$74,999   vs. Less than \$35,000   & $\beta_7$   &  0.25 (0.15, 0.34) & 1.284 \\ 
\$75,000 to  \$149,999   vs.  Less than \$35,000 & $\beta_8$   & 0.43 (0.33, 0.53) & 1.537 \\ 
 \$150,000 or above vs.  Less than \$35,000      &  $\beta_9$   & 0.77 (0.64, 0.89) & 2.160 \\

& & & & \\
Gender: & & & \\
Female vs. Male & $\beta_{10}$  & 0.04 (-0.03, 0.11) & 1.041 \\ 

\hline 
\end{tabular}
 \vspace*{10pt}
\end{table*}


The fitted model says that by holding race, income, and gender at a fixed value, the odds of taking COVID-19 vaccines for a person with an associate's degree over the odds of taking COVID-19 vaccines for a person with a high school graduate or less is 1.57. That is, a person with an associate's degree is more likely to take the vaccines (the odds are 57\% higher) compared to a person with a high school graduate or less. These numbers significantly increase for a person with a bachelor's degree and a graduate degree. We find that That is, the odds of taking the vaccine is around 3.46 times higher for a person with bachelor's degree and is around 6.00 times higher for a person with a graduate degree, both compared to a person with a high school graduate or less.

For the variable race in the model, black and Asian individuals are more likely to take the vaccine compared to white individuals. The odds of taking the vaccine is around 50\% higher in the black population, and around 150\% higher in the Asian population, both are compared to the white population. However, individuals from other races (including race in combination) are less likely to take the vaccines compared to white individuals. We find that the odds of taking the vaccine in other races are 36\% less than the odds in white individuals. 

The coefficients for income show that a person with a higher household income per year (\$35,000 to  \$74,999, or \$75,000 to  \$149,999, or \$150,000 or above) is more likely to take the vaccine compared to a person with less than \$35,000 household income per year. The odds are around 30\%, 50\%, and 100\% more for these three income groups, respectively, compared to the income group with less than \$35,000 household income per year. 

Finally, for the variable gender, we see that the 95\% credible interval of the estimated parameter contains zero, and the odds ratio is approximately 1 (1.04). Therefore, we can say that there is no significant difference in taking COVID-19 vaccines between male and female individuals. 

\begin{figure*}[!ht]
     \centering
       \includegraphics[width=1.0\textwidth]{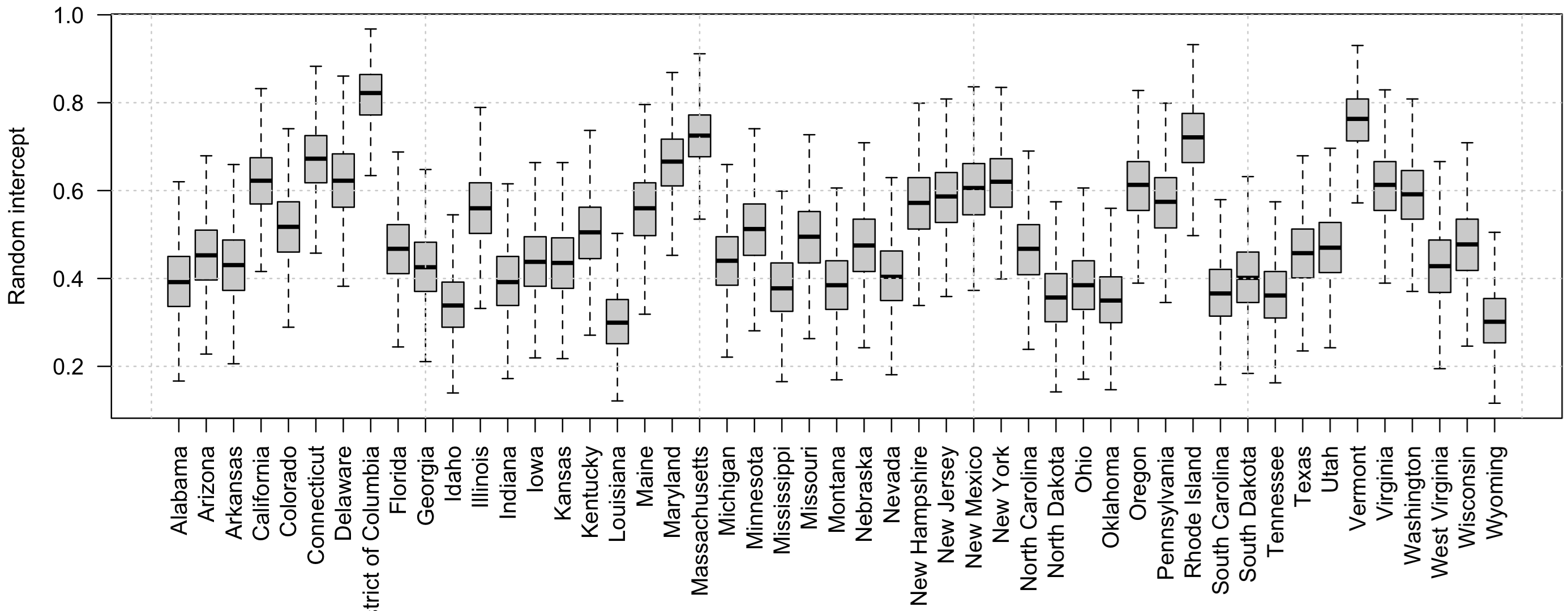}
	\caption{State-wise random intercept.}
        \label{fig:Varying_intercept}
\end{figure*}

The fixed intercept of the model is -0.74, which represents the overall log-odds of taking the vaccine when there are no predictors. Here $\text{P}(Y_i = 1)=\exp{(-0.74)}/(1+\exp{(-0.74)})=0.323$. That is, people have an average 32.3\% chance of taking the COVID-19 vaccines when there are no covariates. The state-wise intercepts (random effect) are shown in Figure~\ref{fig:Varying_intercept}. We find that the District of Columbia, and the states of Massachusetts, Maryland, Rhode Island, and Vermont have high random intercepts (consequently, high overall intercepts). That is, these states have high overall COVID-19 vaccine uptakes. This result supports the descriptive analysis in Figure~\ref{fig:Series_Complete_18PlusPop_Pct}.

\begin{figure*}[ht]
     \centering
       \includegraphics[width=0.85\textwidth]{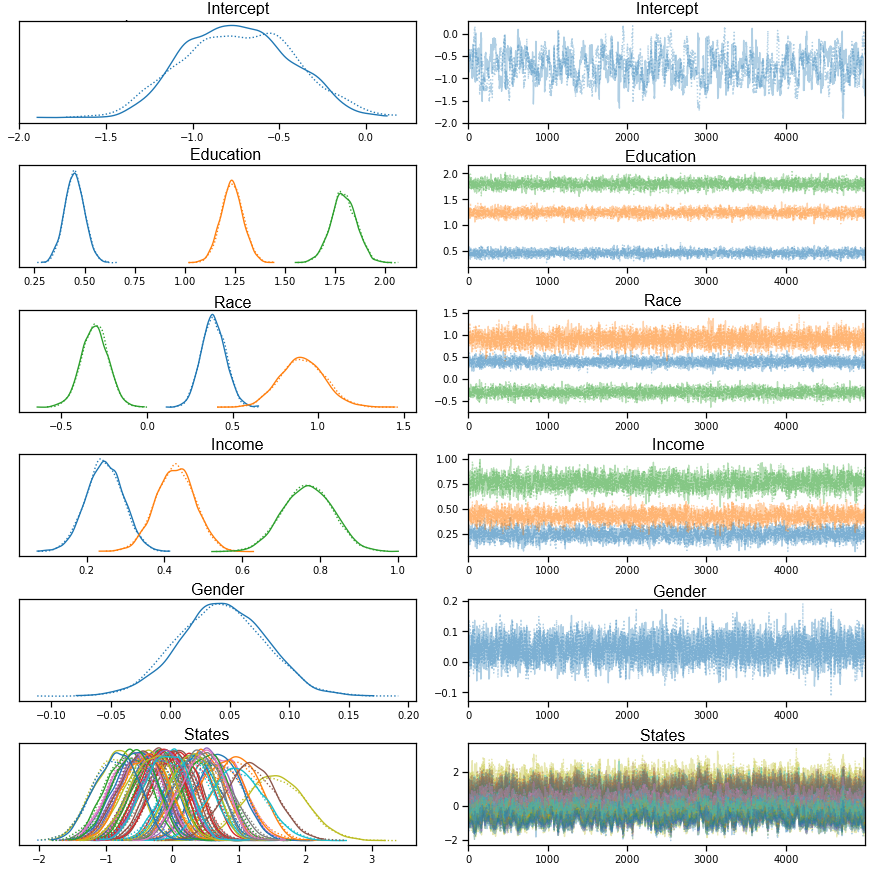}
	\caption{Bayesian Convergence - density plots and trace plots of the parameters.}
        \label{fig:Bayesian_Convergence}
\end{figure*}

Finally, Figure~\ref{fig:Bayesian_Convergence} represents MCMC diagnostic plots for the fixed effect and random effect parameters of the logistic regression parameters. The density plots and trace plots ensure the parameter convergences.

\section{Conclusion}
\label{sec:Conclusion}

In this study, we assess different socio-demographic and geographical determinants of the COVID-19 vaccine uptake in the US based on hierarchical clustering and Bayesian mixed-effects modeling. We use the Household Pulse Survey (HPS) data from the US Census Bureau and the US county-level COVID-19 vaccination dataset from the Centers for Disease Control and Prevention (CDC). The clustering algorithm shows that there are three main groups in the US states according to their COVID-19 vaccination rates. 

 The fitted mixed-effects model provides the probabilistic inference about the vaccine acceptance determinants with uncertainty quantification. The outputs of the Bayesian mixed-effects model show that education and high household income significantly increase COVID-19 vaccine uptake. We also find that African American and Asian individuals have 
higher probabilities of accepting COVID-19 vaccines compared to white individuals. However, there is no significant difference in taking COVID-19 vaccines between male and female individuals.

\section*{Contributions to knowledge}

\noindent What does this study add to existing knowledge?

\begin{itemize}
    \item The current studies on the determinants of people’s COVID-19 vaccine acceptance in the US pay limited attention to the uncertainties of the coefficients of the factors of the COVID‑19 vaccine acceptance.

 \item This is the first study that uses a Bayesian mixed-effects model for the probabilistic inference of the determinants of COVID‑19 vaccine acceptance and their uncertainty quantification. 
    
\item This study conducts spatial clustering of COVID‑19 vaccine uptake in the US and compares distributions of the COVID‑19 vaccine acceptance in different clusters. 
\end{itemize}

\noindent What are the key implications for public health interventions, practice, or policy?
\begin{itemize}
    \item The results will be used to inform interventions for future vaccine rollouts in the US.
    
    \item This study revealed inequalities in COVID‑19 vaccine acceptance in different races, income, and education groups. Public health interventions may need to target particular racial categories and socio-demographic groups, in particular, persons with lower household income and less education.

    \item Interventions also need to target the South, Midwest (except the Upper Midwest), and Mountain West regions, where the vaccine acceptance rate is the lowest.

\end{itemize}

\backmatter




\section*{Declarations}
\begin{itemize}
\item Conflict of interest: The author declares no conflict of interest.
\item Ethics approval and consent to participate: Not applicable
\item Consent for publication: Not applicable
\item Data availability: The data sets are publicly available through the US Census Bureau and the Centers for Disease Control and Prevention (CDC).
\item Code availability: All codes are available upon request. 
\item Author contribution: Not applicable
\end{itemize}









\bibliography{Mixed_effect,sim,network}

\end{document}